\documentclass[aps,pra,twocolumn,showpacs,superscriptaddress ]{revtex4-1}
\usepackage{enumerate}
\usepackage{graphicx}
\usepackage{amsmath}
\usepackage{amsfonts}
\usepackage{amssymb}
\usepackage{epstopdf}
\usepackage{braket}
\usepackage{xcolor}

\begin{document}
\today
\title{Single-loop multiple-pulse nonadiabatic holonomic quantum gates}
\author{Emmi Herterich}
\affiliation{Department of Physics and Astronomy, Uppsala University, Box 516, 
Se-751 20 Uppsala, Sweden}
\author{Erik Sj\"oqvist}
\email{e-mail: erik.sjoqvist@physics.uu.se} 
\affiliation{Department of Physics and Astronomy, Uppsala University, Box 516, 
Se-751 20 Uppsala, Sweden}
\begin{abstract} 
Nonadiabatic holonomic quantum computation provides the means to perform fast and robust 
quantum gates by utilizing the resilience of non-Abelian geometric phases to fluctuations 
of the path in state space. While the original scheme [New J. Phys. {\bf 14}, 103035  
(2012)] needs two loops in the Grassmann manifold (i.e., the space of computational subspaces 
of the full state  space) to generate an arbitrary holonomic one-qubit gate, we propose  
single-loop one-qubit gates that constitute an efficient universal set of holonomic gates 
when combined with an entangling holonomic two-qubit gate. Our one-qubit gate is realized by 
dividing the loop into path segments, each of which is generated by a $\Lambda$-type Hamiltonian. 
We demonstrate that two path segments are sufficient to realize arbitrary single-loop holonomic 
one-qubit gates. We describe how our scheme can be implemented experimentally in a generic atomic 
system exhibiting a three-level $\Lambda$-coupling structure, by utilizing carefully chosen laser 
pulses. 
\end{abstract}
\maketitle
\section{Introduction} 
Holonomic quantum computation (HQC) is the idea that quantum information processing 
can be performed by means of non-Abelian geometric phases. It was first proposed 
\cite{zanardi99} for adiabatic holonomies \cite{wilczek84}, and subsequently generalized 
\cite{sjoqvist12} to nonadiabatic non-Abelian geometric phases \cite{anandan88}. An 
important feature of HQC is the inherent robustness of geometric phases under fluctuations 
of the path in state space \cite{pachos01,solinas12}. 

A key ingredient of HQC is the removal of dynamical phase effects during the execution of 
quantum gates. In the nonadiabatic case, which is the focus of the present paper, this is achieved 
in a three-level $\Lambda$ system, where the two levels encoding a qubit are coupled to an 
excited state by external field pulses. Nonadiabatic HQC in this configuration has been realized 
experimentally for a superconducting artificial atom \cite{abdumalikov13}, NMR \cite{feng13}, 
and NV-centers in diamond \cite{arroyo14,zu14}. The $\Lambda$-system-based HQC has been 
combined with decoherence free subspaces \cite{xu12,xu14a,liang14,xue15,zhou15,xue16}, 
noiseless subsystems \cite{zhang14}, and dynamical decoupling \cite{xu14b}. The nonadiabatic 
property makes it possible to shorten the exposure to undesired external influences 
\cite{sjoqvist12,johansson12}. 

The essential geometric structure of nonadiabatic HQC is the complex Grassmann manifold 
$\mathcal{G} (N;K)$, i.e., the space of $K$-dimensional subspaces of an $N$-dimensional 
state space \cite{bengtsson06}. A loop in the Grassmannian generates a holonomic quantum 
gate acting on the target computational subspace encoded at the common start- and end-point. 

The $\Lambda$-system-based holonomic gates in Ref.~\cite{sjoqvist12} utilize resonant 
laser pulses. Here, two distinct loops in the corresponding Grassmannian 
$\mathcal{G} (3;2)$ are needed to perform an arbitrary holonomic one-qubit gate. Experimentally, 
the two loops correspond to two consecutive laser pulse pairs of arbitrary shape.

The need for two loops in order to implement an arbitrary holonomic one-qubit gate is an 
apparent drawback as it doubles the exposure time to various error sources. This motivates 
attempts to try to reduce the number of loops. It has recently been shown \cite{xu15,sjoqvist16} 
that an arbitrary one-qubit gate can be achieved for a single loop by using off-resonant laser 
pulses. However, this off-resonant scheme has two disadvantages. First, it requires 
square pulses, a restriction that blocks the possibility to optimize robustness by tailoring 
the pulse shape; secondly, the small rotation angle limit would correspond to either very short 
pulses or to small field amplitudes, both of which would lead to unstable gate operations.

Here, we demonstrate that these problems can be resolved. To this end, we propose a  
single-loop multiple-pulse scheme, in which the loop is divided into segments. Our scheme is 
conceptually akin to the `orange slice' path on the Bloch sphere commonly used when observing 
the Abelian geometric phase in quantum optics experiments \cite{kwiat91,allmann97,du03,rippe08}. 
We demonstrate that our proposed scheme is able to perform arbitrary holonomic 
one-qubit gates for fewer loops in the Grassmannian than in the original scheme 
\cite{sjoqvist12}, while keeping the full flexibility concerning the choice of 
laser pulse shape and pulse duration. Universal HQC can be achieved by combining 
our holonomic one-qubit gate with an entangling holonomic 
two-qubit gate (e.g., \cite{sjoqvist12,gurkan15}).  

The outline of the paper is as follows. The next section reviews earlier versions of 
$\Lambda$-system-based holonomic gates. Section \ref{sec:iterative} outlines our 
single-loop multiple-pulse scheme, first by describing the general idea, and thereafter 
followed by demonstrating that an arbitrary holonomic one-qubit  gate can be realized 
by dividing the loop into just two path segments. In Sec.~\ref{sec:experimental}, we delineate 
how our scheme can be implemented experimentally. The paper ends with the conclusions. 

\section{Holonomic gates in the $\Lambda$-system}
In the $\Lambda$-configuration, a laser pulse pair induces transitions between the qubit 
states $\ket{0}$, $\ket{1}$ and the excited state $\ket{e}$ of a generic three-level system. 
This is described by the Hamiltonian (we put $\hbar = 1$ from now on) 
\begin{eqnarray} 
\label{eq:hamiltonian1_01}
\mathcal{H}(t) & = & \Delta_0\ket{0}\bra{0}+\Delta_1\ket{1}\bra{1}
\nonumber\\
 & & + \Upsilon_0 (t) \ket{e} \bra{0} + 
\Upsilon_1 (t) \ket{e} \bra{1} + \text{h.c.} ,
\end{eqnarray}
where we have used the rotating wave approximation in the interaction picture. The 
complex-valued ratio $\Upsilon_0 (t)/\Upsilon_1 (t)$ describes the relative amplitude 
and phase between the laser pulses; $\Delta_p$, $p=0,1$, are detunings. 

\begin{figure*}[ht]
\centering
\includegraphics[scale=.6]{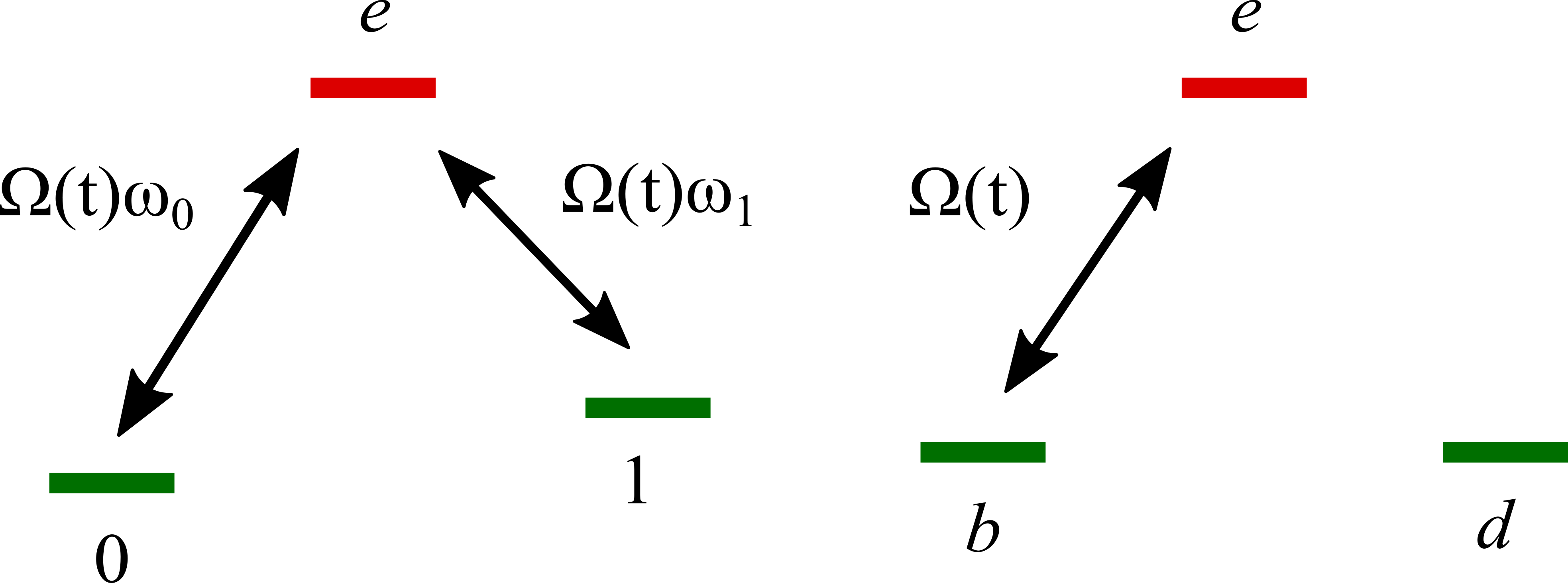}
\caption{The $\Lambda$-system. In the left panel, we see how the qubit states $\ket{0}$ 
and $\ket{1}$, defining our target computational subspace, are controlled by the laser 
parameters $\Omega(t) \omega_0$ and $\Omega(t)\omega_1$, respectively. The dynamics 
can be understood as a Rabi oscillations between the bright state $\ket{b} = 
\omega_0^{\ast} \ket{0} + \omega_1^{\ast} \ket{1}$ and excited state $\ket{e}$, while 
the dark state $\ket{d} = -\omega_1 \ket{0} + \omega_0 \ket{1}$ decouples from the 
system, as shown in the right panel.}
\label{fig:lambdasystems}
\end{figure*}

Holonomic quantum information processing in the $\Lambda$-system is implemented 
by applying the laser pulses simultaneously and on resonance \cite{sjoqvist12}. In other 
words, $\Upsilon_p (t) = \Omega (t) \omega_p$ and $\Delta_p=0$. Here, $\Omega (t)$ 
is real-valued and has nonvanishing support over the duration $\tau$ of the pulse pair. 
The time-independent $\omega_p$ are assumed to satisfy the normalization relation 
$|\omega_0|^2 + |\omega_1|^2=1$. 

To see how these parameter choices implement a purely holonomic gate acting on 
$\textrm{Span} \{ \ket{0},\ket{1} \}$, it is convenient first to express the Hamiltonian in 
terms of the dark and bright states $\ket{d} = -\omega_1\ket{0} + \omega_0\ket{1}$ and 
$\ket{b} = \omega_0^{\ast} \ket{0} + \omega_1^{\ast} \ket{1}$, respectively. One thereby 
finds  
\begin{eqnarray} 
\label{eq:hamiltonian}
\mathcal{H}(t) = \Omega(t)\big( \ket{e} \bra{b} + \ket{b} \bra{e} \big) \equiv \Omega (t) H,  
\label{eq:dbhamiltonian}
\end{eqnarray} 
which shows that the evolution can be understood as a Rabi oscillations between $\ket{b}$ 
and $\ket{e}$ with frequency $\Omega (t)$, while $\ket{d}$ decouples from the system. The 
$\Lambda$-configuration in the $\ket{0},\ket{1}$ and $\ket{d},\ket{b}$ representations 
is shown in Fig.~\ref{fig:lambdasystems}. The Hamiltonian in Eq.~(\ref{eq:dbhamiltonian}) 
moves the qubit subspace $\textrm{Span} \{ \ket{0},\ket{1} \}$ in the full state space 
$\textrm{Span} \{ \ket{0},\ket{1},\ket{e} \}$; a process that can be viewed as a path in 
the Grassmannian $\mathcal{G} (3;2)$. Each point along the path in $\mathcal{G}(3;2)$ 
is spanned by the vectors 
\begin{eqnarray}
\ket{\psi_d(a)} & = & \mathcal{U}(a,0) \ket{d} = \ket{d} , 
\nonumber\\
\ket{\psi_b(a)} & = & \mathcal{U}(a,0)\ket{b} = \cos(a)\ket{b}-i\sin(a)\ket{e} , 
\end{eqnarray}
where
\begin{eqnarray} 
\label{eq:laserpulsearea}
a=\int_{0}^{t}\Omega(t')dt' 
\end{eqnarray}
is the pulse area and $\mathcal{U}(a,0) = \exp(-iaH)$ is the time evolution operator. A full 
loop $C_{\bf n}$ in the Grassmannian is realized when $a \equiv a_1 =\pi$. The transformation 
on the one-qubit subspace is purely holonomic (i.e., depends only on $C_{\bf n}$) as the 
dynamical matrix elements $\bra{\psi_k (a)} \mathcal{H}(t) \ket{\psi_l (a)}$, with $k,l=b,d$, all 
vanish for $a\in [0,\pi]$. Explicitly, one finds \cite{sjoqvist12,kult06} 
\begin{eqnarray} 
U(C_{\bf n}) = \mathcal{U}(\pi,0) \mathbb{P}(0) = ie^{-i\frac{1}{2} \pi {\bf n} \cdot 
\boldsymbol{\sigma}} = {\bf n} \cdot \boldsymbol{\sigma}, 
\label{eq:1pulse}
\end{eqnarray}
where ${\bf n} = ( \sin\theta\cos\phi, \sin\theta\sin\phi, \cos\theta )$ is a unit vector 
defined by $\omega_0/\omega_1 = -e^{i\phi} \tan \frac{\theta}{2}$, $\mathbb{P}(0) = 
\ket{d}\bra{d} + \ket{b}\bra{b} = \ket{0}\bra{0} + \ket{1}\bra{1}$ is the projection operator 
onto the target computational subspace encoding the qubit, and $\boldsymbol{\sigma} = 
(\sigma_x,\sigma_y,\sigma_z)$ are the standard Pauli operators expressed in the 
$\ket{0},\ket{1}$ basis. The resulting unitary transformation $U(C_{\bf n})$ is the holonomic 
one-qubit gate associated with the loop $C_{\bf n}$. 

The holonomy in Eq.~(\ref{eq:1pulse}) shows that a single pulse pair can generate only traceless 
one-qubit gates. To achieve arbitrary holonomic gates, it is necessary to apply two consecutive 
laser pulse pairs, each with pulse area $\pi$, which corresponds to traversing two loops in the 
Grassmannian. To see this, assume that the two pulse pairs generate loops $C_{{\bf n}_1}$ 
and $C_{{\bf n}_2}$, characterized by laser parameters that correspond to unit vectors ${\bf n}_1$ 
and ${\bf n}_2$, the resulting composite holonomy transformation becomes 
\begin{eqnarray}
U(C) & = & U(C_{{\bf n}_2}) U(C_{{\bf n}_1}) 
\nonumber \\ 
 & = & {\bf n}_1 \cdot {\bf n}_2 \ \mathbb{P}(0) - i ({\bf n}_1 \times {\bf n}_2) \cdot 
\boldsymbol{\sigma} . 
\label{eq:twoloop}
\end{eqnarray} 
This is an arbitrary SU(2) transformation that rotates the qubit by an angle $2 \arccos 
({\bf n}_1\cdot {\bf n}_2)$ around the normal of the plane spanned by ${\bf n}_1$ and ${\bf n}_2$.  

The need for two loops is an apparent drawback as it doubles the exposure time to various 
error sources. Thus, it is desirable to find methods that can realized holonomic one-qubit gates 
for a single loop in the Grassmannian. It has recently been shown \cite{xu15,sjoqvist16} that 
off-resonant, equally detuned laser pulses can be used to implement arbitrary single-loop 
holonomic one-qubit gates. This is described by the Hamiltonian 
\begin{eqnarray}
\mathcal{H}_{\Delta} (t) = \Delta \ket{e} \bra{e} + \Omega(t)\big( \ket{e} \bra{b} + \ket{b} \bra{e} \big) 
\end{eqnarray}
with a trivial shift of the zero-point energy and $\Delta$ being the detuning \cite{remark1}. 
In order to preserve the geometric character of the evolution, the Hamiltonian needs to 
commute with itself during the pulse, which implies that $\Omega (t)$ must be square-shaped, 
i.e., $\Omega (t) = \Omega_0$ for $0 \leq t \leq \tau$ and zero otherwise. The evolution 
becomes cyclic corresponding to a loop $C_{{\bf n};\Delta}$ in the Grassmannian if 
\begin{eqnarray}
\tau = \frac{2\pi}{\sqrt{\Delta^2 + 4\Omega_0^2}} .
\end{eqnarray}
One finds 
\begin{eqnarray} 
U(C_{{\bf n};\Delta}) = e^{i\frac{1}{2} (\pi - \chi)} 
e^{-i\frac{1}{2} (\pi - \chi) {\bf n} \cdot \boldsymbol{\sigma}} ,
\label{eq:1pulse_univ}
\end{eqnarray}
where 
\begin{eqnarray}
\chi = \frac{\pi\Delta}{\sqrt{\Delta^2 + 4\Omega_0^2}} . 
\end{eqnarray}
The gate $U(C_{{\bf n};\Delta})$ is an arbitrary holonomic one-qubit gate as the rotation angle 
$\pi-\chi$ can be varied between zero and $\pi$ by decreasing $\Delta /(2\Omega_0)$ from 
infinity to zero. As a consistency check, we may note that $U(C_{{\bf n};\Delta})$ reduces to 
$U(C_{\bf n})$ in the $\Delta /(2\Omega_0) \rightarrow 0$ limit. 

Although $U(C_{{\bf n};\Delta})$ covers all one-qubit gates, it suffers from two disadvantages.  
First, if $\Delta \neq 0$, then the pulse must be square-shaped in order to preserve the 
geometric character of the gate, which is a practical limitation as full shape flexibility 
is an important feature needed to optimize robustness to different kinds of errors (see, e.g., 
\cite{roos04}). Secondly, the small rotation angle limit is achieved for large $\Delta/(2\Omega_0)$.  
This can be reached either by using a large $\Delta$ and thereby a small $\tau$, which makes the 
gate highly unstable to small perturbations in the run-time \cite{spiegelberg13}, or by using 
a small $\Omega_0$, which introduces an instability similar to fluctuations in the field amplitude. 
In the following section, we demonstrate a multiple-pulse method to realize arbitrary 
single-loop holonomic one-qubit gates, which avoids these disadvantages. 

\section{Single-loop multiple-pulse scheme}
\label{sec:iterative}
\subsection{General setting}
\begin{figure*}[ht]
\centering
\includegraphics[scale=0.37]{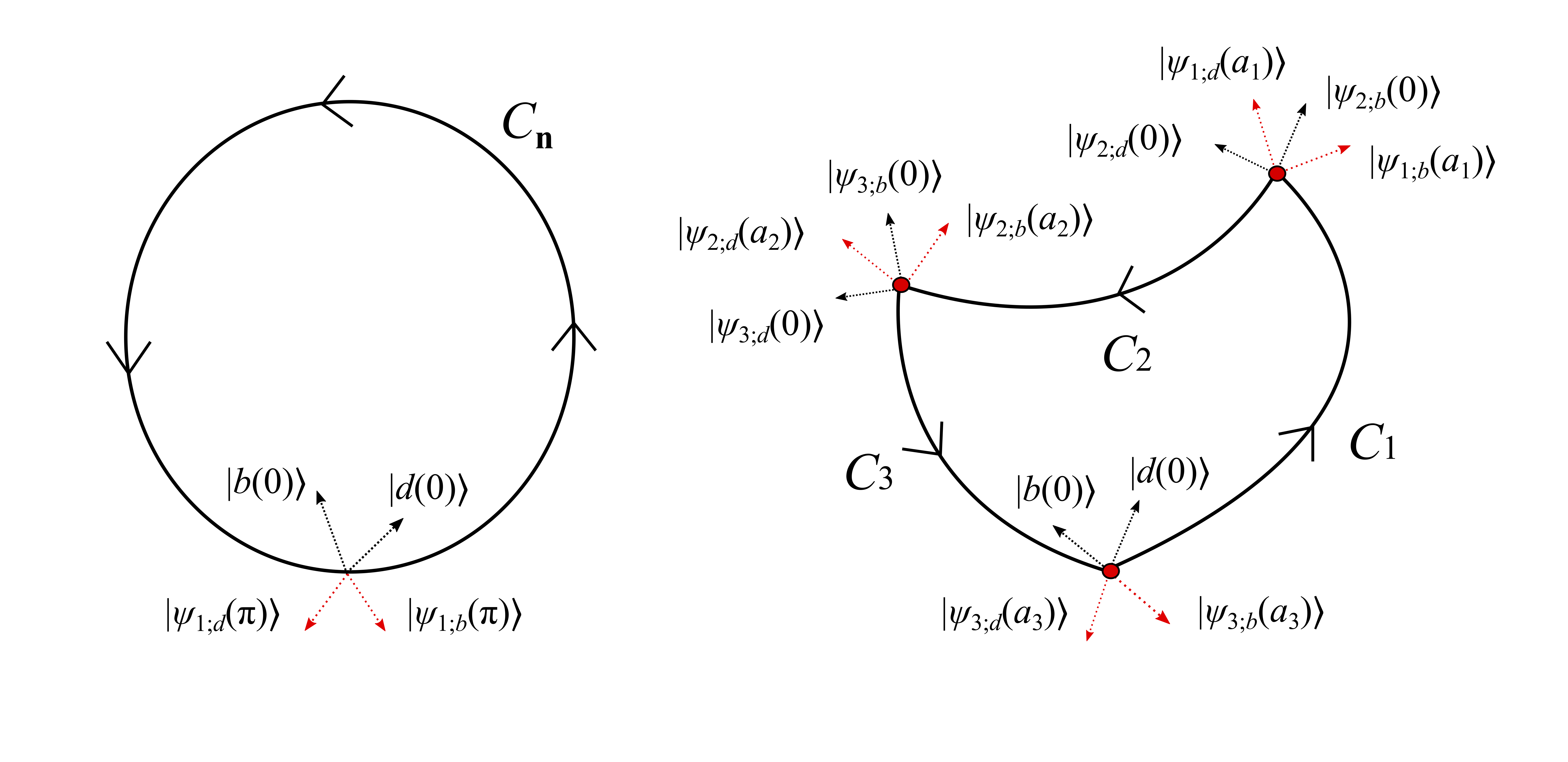}
\caption{The single-loop multiple-pulse scheme in the $\Lambda$ system. The left panel 
shows the case of a loop generated by a single pulse pair. The initial point in the Grassmannian 
$\mathcal{G}(3;2)$ spanned by the vectors $\ket{k}$, $k=b,d$, where $\ket{b} = 
\omega_0^{\ast} \ket{0} + \omega_1^{\ast} \ket{1}$ and $\ket{d} = -\omega_1 \ket{0} + 
\omega_0 \ket{1}$, makes one full revolution by following the path generated by the $\Lambda$ 
Hamiltonian $\mathcal{H}(t) = \Omega (t) \left( \ket{e} \bra{b} + \ket{b} \bra{e} \right)$. 
This induces the holonomic one-qubit transformation $\ket{k} \mapsto \ket{\psi_k} = 
U(C_{\bf n}) \ket{k}$, ${\bf n}$ being determined by the laser parameters $\omega_0$ and 
$\omega_1$. The closing of the path is ensured by choosing pulse area $a \equiv a_1 = 
\int_0^{\tau} \Omega (t) dt = \pi$. The right panel visualizes the multiple-pulse scheme, in which 
the loop is divided into path segments. Here, the initial subspace moves along the first path 
segment $C_1$ under $\mathcal{H}_1(t) = \mathcal{H}(t)$ to a point spanned by $\ket{\psi_{1;k}(a_1)}$ 
by chosing pulse area $a_1\neq \pi$.  A unitary transformation $\ket{\psi_{1;k}(a_1)} 
\mapsto \ket{\psi_{2;k}(0)}$ and $\ket{\psi_{1;e}(a_1)} \mapsto \ket{\psi_{2;e}(0)} = 
\ket{\psi_{1;e}(a_1)}$, defines a new $\Lambda$ Hamiltonian $\mathcal{H}_2(t) = \Omega_2 (t) 
\left( \ket{\psi_{2;e}(0)} \bra{\psi_{2;b}(0)}  + \ket{\psi_{2;b}(0)} \bra{\psi_{2;e}(0)} \right)$ that 
generates the second path segment $C_2$. This procedure is repeated $L$ times (here, the 
$L=3$ case is shown). If the final point of the $L$th segment coincides with $\textrm{Span} 
\{ \ket{b},\ket{d} \} = \textrm{Span} \{ \ket{0},\ket{1} \}$, then $C_1 \ast \cdots \ast C_L$ 
forms a loop. In this case, the resulting transformation $\ket{k} \mapsto \ket{\psi_{L;k} (a_L)} = 
U(C_1 \ast \cdots \ast C_L) \ket{k}$ is unitary and constitutes our holonomic single-loop 
multiple-pulse one-qubit gate.}
\label{fig:one_loop}
\end{figure*}

Consider a path in $\mathcal{G} (3;2)$ divided into $L$ segments $C_1,\ldots,C_L$, generated 
by $L$ pulse pairs with pulse areas $a_1,\ldots,a_L$.  Figure \ref{fig:one_loop} schematically depicts 
such a division when $C_1 \ast \cdots \ast C_L$ is a loop and $L=3$. The process of dividing 
the path can be described as the following iterative procedure: 
\begin{itemize}
\item[(i)] The first path segment starts at the target computational subspace $\textrm{Span} 
\{ \ket{b},\ket{d} \} = \textrm{Span} \{ \ket{0},\ket{1} \}$ and is generated by the zero-detuned 
(resonant) Hamiltonian $\mathcal{H}_1 (t)$, being identical to $\mathcal{H}(t)$ in 
Eq.~(\ref{eq:hamiltonian}). 
\item[(ii)] The initial point $\textrm{Span} \{ \ket{\psi_{n;b} (0)},\ket{\psi_{n;d} (0)} \}$ of the 
$n$th path segment coincides with the final point $\textrm{Span} \{ \ket{\psi_{n-1;b} (a_{n-1})}, 
\ket{\psi_{n-1;d} (a_{n-1})} \}$ of the $(n-1)$th path segment, for $n=2,\ldots, L$.  
\item[(iii)] The resonant Hamiltonian driving the evolution along the $n$th path segment reads 
\begin{eqnarray} 
\mathcal{H}_n (t) & = & \Omega_n(t) \big( \ket{\psi_{n;e}(0)}\bra{\psi_{n;b}(0)} 
\nonumber \\ 
 & & + \ket{\psi_{n;b}(0)}\bra{\psi_{n;e}(0)}\big) \equiv \Omega_n (t) H_n , 
\label{eq:hamiltonian_gen} 
\end{eqnarray}
where   
\begin{eqnarray} 
\ket{\psi_{n;k}(0)} & = & V_n \ket{\psi_{n-1;k}(a_{n-1})}, \quad  k=b,d,  
\label{eq:initialstatesgen}
\end{eqnarray}
and $a_{n-1} = \int_0^{\tau_{n-1}} \Omega_{n-1}(t) dt$ with $\tau_{n-1}$ being the corresponding 
run-time. Here, the basis transformation $V_n$ acts unitarily on the final subspace of the $(n-1)$th 
segment, which further implies that $V_n \ket{\psi_{n-1;e}(a_{n-1})} = \ket{\psi_{n-1;e}(a_{n-1})}$. 
Physically, $V_n$ defines the discrete changes of the external laser fields when moving 
from $C_{n-1}$ to $C_n$.
\end{itemize}

The time-evolution operator along the $n$th path segment evaluated at pulse area $a_n$, 
takes the form \cite{johansson12}
\begin{eqnarray} 
\mathcal{U}_n(a_n,0) & = & e^{-ia_n H_n} =  \ket{\psi_{n;d}(0)}\bra{\psi_{n;d}(0)} 
\nonumber\\
 & & + \cos a_n \big( \hat{1} - \ket{\psi_{n;d}(0)}\bra{\psi_{n;d}(0)} \big) 
\nonumber\\
 & & -i\sin a_n \big(\ket{\psi_{n;e}(0)}
\bra{\psi_{n;b}(0)} 
\nonumber\\
 & & + \ket{\psi_{n;b}(0)}\bra{\psi_{n;e}(0)}\big) . 
\label{eq:timeevop} 
\end{eqnarray}
Due to the $\Lambda$ structure of $H_n$, it follows that the evolution of the computational 
subspace is purely geometric along all path segments. We thus find the holonomy \cite{kult06}
\begin{eqnarray}
U(C_1 \ast \cdots \ast C_L) & = & 
\mathcal{U}_L (a_L,0) \cdots \mathcal{U}_1 (a_1,0) \mathbb{P}(0) . 
\end{eqnarray}
By carefully choosing laser parameters so that $C_1 \ast \cdots \ast C_L$ forms a loop, 
$U(C_1 \ast \cdots \ast C_L)$ is a unitary operator acting on $\textrm{Span} \{ \ket{0},\ket{1} \}$. 
In such a case, $U(C_1 \ast \cdots \ast C_L)$ is our one-qubit gate. 

\subsection{$L=2$ holonomic gates}
We now demonstrate that two pulse pairs ($L=2$) with $a_1=a_2=\pi/2$ are sufficient to 
construct an arbitrary holonomic one-qubit quantum gate by traversing a single loop in 
$\mathcal{G} (3;2)$. 

Our starting point is $\ket{\psi_{1;e}(0)} =\ket{e}$, $\ket{\psi_{1;b}(0)} = \ket{b}$, and 
$\ket{\psi_{1;d}(0)}=\ket{d}$, where the two latter vectors span the target computational 
subspace. By directly evaluating the time evolution operator in Eq.~(\ref{eq:timeevop}) at 
$a_1 = \pi/2$, we obtain
\begin{eqnarray}
\mathcal{U}_1\left( \pi/2,0\right)	
=\ket{d}\bra{d}-i\big(\ket{e}\bra{b}+\ket{b}\bra{e}\big) , 
\label{eq:1stunitary}
\end{eqnarray}
which yields 
\begin{eqnarray} 
\label{eq:timeevstatea1}
\ket{\psi_{1;e}\left( \pi/2 \right)} & = &
\mathcal{U}_1\left( \pi/2,0\right)\ket{e} = -i\ket{b} ,  
\nonumber \\
\ket{\psi_{1;b}\left( \pi/2 \right)} & = & 
\mathcal{U}_1\left( \pi/2,0\right)\ket{b} = -i\ket{e} , 
\nonumber \\
\ket{\psi_{1;d}\left( \pi/2 \right)} & = & 
\mathcal{U}_1\left( \pi/2,0\right)\ket{d} = \ket{d}.
\end{eqnarray}
The next step is to find the vectors $\ket{\psi_{2,k}(0)}$ spanning the initial point of 
the second path segment $C_2$. We can make a nontrivial choice of these vectors so that 
the final point of $C_2$ coincides with the initial point of $C_1$, i.e., so that $C_1 \ast C_2$ 
forms a loop. The choice is   
\begin{eqnarray}
\ket{\psi_{2;e}(0)} &= & V_2 \ket{\psi_{1;e}\left( \pi/2 \right)} = -i\ket{b}, 
\nonumber \\
\ket{\psi_{2;b}(0)} &= & V_2 \ket{\psi_{1;b}\left( \pi/2 \right)} =-ie^{i\eta}\ket{e},
\nonumber\\
\ket{\psi_{2;d}(0)} & = & V_2 \ket{\psi_{1;d}\left( \pi/2 \right)} = e^{-i\eta}\ket{d} , 
\label{eq:2ndstatesinitial}
\end{eqnarray}
as given by the basis transformation
\begin{eqnarray} 
\label{eq:rot_op}
V_2 & = & \ket{\psi_{1;e}\left( \pi/2 \right)} \bra{\psi_{1;e}\left( \pi/2 \right)} 
\nonumber \\ 
 & & + e^{i\eta} \ket{\psi_{1;b}\left( \pi/2 \right)} \bra{\psi_{1;b}\left( \pi/2 \right)} 
\nonumber \\ 
 & & + e^{-i\eta} \ket{\psi_{1;d}\left( \pi/2 \right)} \bra{\psi_{1;d}\left( \pi/2 \right)} 
\nonumber \\ 
 & = & \ket{b} \bra{b} + e^{i\eta} \ket{e} \bra{e} + e^{-i\eta} \ket{d} \bra{d} . 
\end{eqnarray}
The resulting Hamiltonian for the second pulse thus reads
\begin{eqnarray} 
\label{eq:2ndhamiltonian}
\mathcal{H}_2(t) & = & \Omega_2(t)\big( e^{-i\eta} \ket{b}\bra{e} + e^{i\eta} \ket{e}\bra{b} \big) 
\nonumber \\ 
 & \equiv & \Omega_2(t) H_2 , 
\end{eqnarray}
which generates the time evolution operator 
\begin{eqnarray}
\mathcal{U}_2\left( \pi/2,0\right) = 
\ket{d}\bra{d}-i\big(e^{-i\eta}\ket{b}\bra{e}+e^{i\eta}\ket{e}\bra{b}\big) 
\label{eq:2ndunitary}
\end{eqnarray} 
when evaluated at $a_2 = \pi/2$. By taking into account the explicit form of the bright state, 
we see that $\mathcal{H}_2(t)$ is equivalent to a shift of the two laser parameters $\omega_p$ 
by the same phase $\eta$, i.e., $\omega_p \mapsto e^{i\eta} \omega_p$. 

Consecutive application of $\mathcal{U}_1\left( \pi/2,0\right)$ and $\mathcal{U}_2 
\left( \pi/2,0\right)$ generates a loop $C_1 \ast C_2$ in the Grassmannian. Thus,   
\begin{eqnarray} 
U(C_1 \ast C_2) = \mathcal{U}_{2}\left( \pi/2,0\right)\,\mathcal{U}_{1}\left( \pi/2,0\right) 
\mathbb{P}(0) , 
\label{eq:sequence}
\end{eqnarray}
is unitary and constitutes the holonomic one-qubit quantum gate. By inserting 
Eqs.~(\ref{eq:1stunitary})  and (\ref{eq:2ndunitary}) into Eq.~(\ref{eq:sequence}), 
we obtain 
\begin{eqnarray}
U(C_1 \ast C_2) & = & \ket{d}\bra{d} - e^{-i\eta} \ket{b}\bra{b} 
\nonumber \\ 
 & = & e^{i\frac{1}{2}(\pi-\eta)} e^{-i\frac{1}{2}(\pi-\eta) {\bf n} \cdot \boldsymbol{\sigma}}.
\label{eq:l2gate}
\end{eqnarray}
The factor $e^{i\frac{1}{2}(\pi-\eta)}$ is a global phase factor that can be ignored. The 
operator $e^{-i\frac{1}{2}(\pi-\eta) {\bf n} \cdot \boldsymbol{\sigma}}$ corresponds to a 
rotation around ${\bf n}$ by an angle $\pi-\eta$, which should be compared to the 
rotation around ${\bf n}_1 \times {\bf n}_2/\left| {\bf n}_1 \times {\bf n}_2 \right|$ by the 
angle $2 \arccos ({\bf n}_1 \cdot {\bf n}_2)$ obtained by traversing two loops in the original 
$\pi$ pulse scheme, as given by Eq.~(\ref{eq:twoloop}). 

\begin{figure*}[ht]
\centering
\includegraphics[scale=.65]{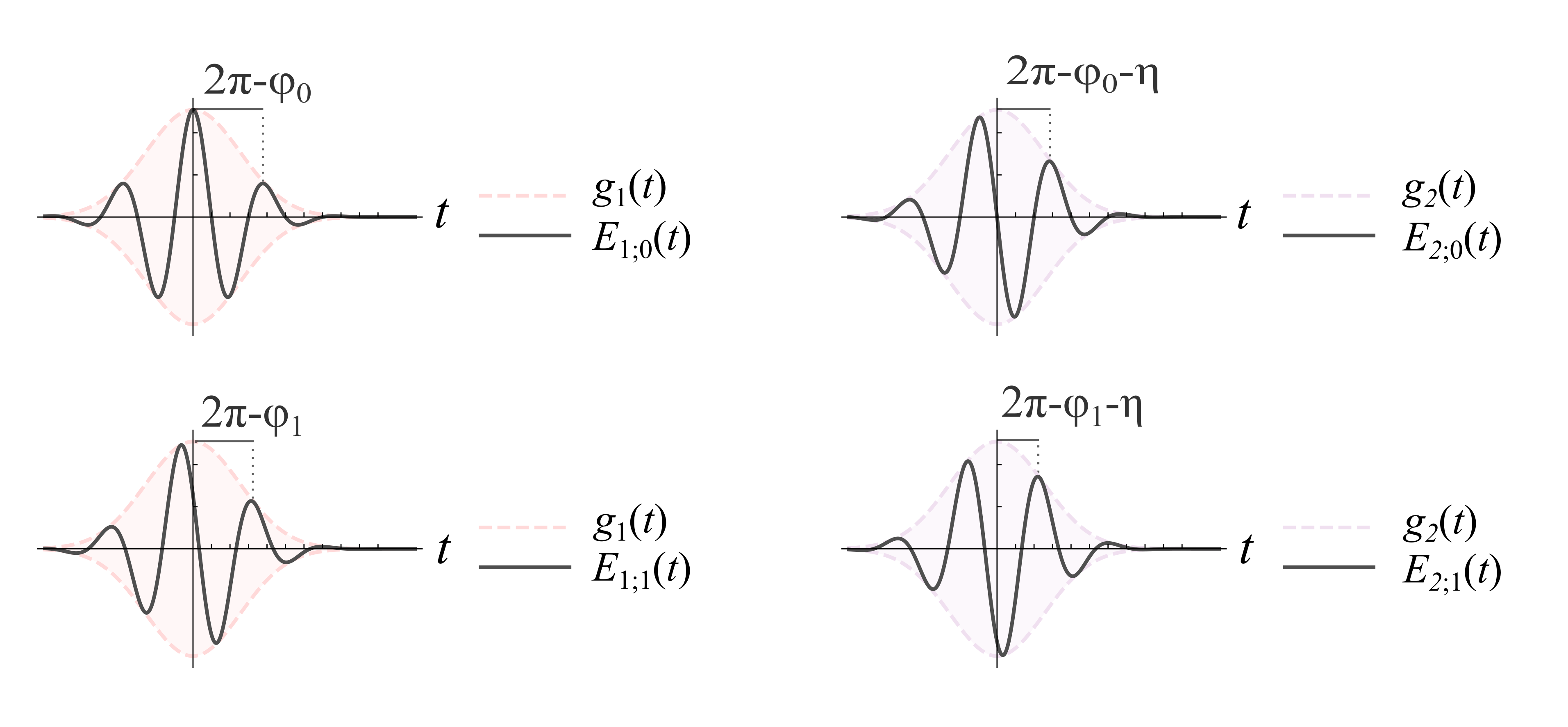}
\caption{Laser pulses that implement $L=2$ holonomic gates. The first (second) pulse pair 
is shown in the left (right) panel. The pulses within each pair are applied simultaneously, while 
the pairs are mutually nonoverlapping in time, but have the same shape. The oscillating solid 
lines are the pulses $E_{1;p}(t) = g_1(t) \cos (f_p t + \varphi_p) \propto |{\bf E}_{1;p} (t)|$ and 
$E_{2;p}(t) = g_2(t) \cos (f_p t + \varphi_p +\eta) \propto |{\bf E}_{2;p} (t)|$, $p=0,1$, 
restricted by the dashed curves $\pm g_1 (t)$ and $\pm g_2 (t)$, respectively. These pulses 
realize an $L=2$ holonomic one-qubit gate provided the area of the envelope functions 
$g_n(t)$, $n=1,2$, is chosen so as to implement $\frac{\pi}{2}$ pulses.}
\label{fig:realization}
\end{figure*}

The holonomic gate $U(C_1 \ast C_2)$ reaches all possible one-qubit transformations by 
separately varying the phase shift  $\eta$ and the laser parameters ${\bf n}$. In contrast to the 
off-resonant scheme proposed in Refs.~\cite{xu15,sjoqvist16}, our gate preserves its geometric 
character for any pulse shape. It is essential in the proposed $L=2$ scheme that the two pulse 
pairs both have area $\pi/2$, in order for the two path segments to form a loop in the 
Grassmannian \cite{remark2}. We further note that the rotation angle $\pi - \eta$ is 
independent of the duration of the pulses, which implies that the small angle limit is achievable 
without violating the rotating wave approximation. Thus, we conclude that our holonomic 
one-qubit gate resolves the problems of the off-resonant scheme \cite{xu15,sjoqvist16}, 
still maintaining the single-loop advantage over the original proposal of Ref.~\cite{sjoqvist12}. 

\section{Experimental implementation}
\label{sec:experimental}
The $L=2$ holonomic gates can be implemented experimentally in electric 
dipole transitions generated by four appropriately phase-shifted laser pulses in a generic 
atomic three-level systems. The four laser pulses should be applied as two consecutive 
pairs, as shown in Fig.~\ref{fig:realization}. The first pair is given by the oscillating electric 
fields ${\bf E}_{1;p} (t) = 
\boldsymbol{\epsilon}_p g_1 (t) \cos (f_p t + \varphi_p)$, $p=0,1$, $g_1(t)$ being the 
envelope function describing the common shape and duration of the pulses. Similarly, 
the second pulse pair is given by ${\bf E}_{2;p} (t) = \boldsymbol{\epsilon}_p g_2 (t) 
\cos (f_p t + \varphi_p + \eta)$ and should not overlap with the first pulse pair (thus, 
$g_1(t)$ and $g_2(t)$ should be mutually nonoverlapping, but have the same shape). 
The polarization $\boldsymbol{\epsilon}_p$ is chosen so as to allow for only the $\ket{p} 
\leftrightarrow \ket{e}$ transition, by utilizing appropriate selection rules. The ratio 
$\left| \boldsymbol{\epsilon}_0 \right|^2/ \left| \boldsymbol{\epsilon}_1 \right|^2$ 
describes the relative intensity of the two laser pulses. We assume that the oscillation 
frequencies $f_p$ are tuned on resonance with the transition frequencies $\nu_{ep}$, 
given by the bare Hamiltonian $\mathcal{H}_{\textrm{bare}} = -\nu_{e0} \ket{0} \bra{0} - 
\nu_{e1} \ket{1} \bra{1}$, for which the energy of the excited state is taken as the 
zero point. 

Now, in the interaction picture, we find  
\begin{eqnarray}
\tilde{H}_1 (t) & = & 
\Omega_1 (t) \left[ \omega_0 \left(1+e^{-2i\nu_{e0}t}\right)\ket{e}\bra{0} \right.  
\nonumber \\ 
 & & \left. + \omega_1 \left(1+e^{-2i\nu_{e1}t}\right)\ket{e}\bra{1} + 
{\textrm{h.c.}} \right] , 
\nonumber \\  
\tilde{H}_2 (t) & = & 
\Omega_2 (t) \left[ \omega_0 \left( e^{i\eta}+e^{-2i\nu_{e0}t -i\eta}\right)\ket{e}\bra{0} \right.  
\nonumber \\ 
 & & \left. + \omega_1 \left( e^{i\eta} + e^{-2i\nu_{e1}t -i\eta}\right)\ket{e}\bra{1} + 
{\textrm{h.c.}} \right] .  
\end{eqnarray}
Here, $\Omega_n (t) \omega_p = e^{i\varphi_p} \bra{e} \boldsymbol{\mu} \cdot 
\boldsymbol{\epsilon}_p \ket{p} g_n (t)/2$, with $n=1,2$ and  $\boldsymbol{\mu}$ 
being the electric dipole operator, which determine the polar angles 
$\theta$ and $\phi$ of ${\bf n}$ according to 
\begin{eqnarray} 
e^{i(\varphi_0 - \varphi_1)} \frac{\bra{e} \boldsymbol{\mu} \cdot 
\boldsymbol{\epsilon}_0 \ket{0}}{\bra{e} \boldsymbol{\mu} \cdot 
\boldsymbol{\epsilon}_1 \ket{1}} = - e^{i\phi} \tan \frac{\theta}{2} .  
\end{eqnarray}
By neglecting the rapidly oscillating terms $e^{\pm 2i\nu_{ep} t}$ and 
$e^{\pm i(2\nu_{ep} t +\eta)}$ (rotating wave approximation), we see that $\tilde{H}_n (t)$  
coincides with $\mathcal{H}_n(t)$, thus demonstrating that the holonomic one-qubit gate in 
Eq.~(\ref{eq:l2gate}) can be realized in this physical setting. 

The superconducting artificial atom experiment in Ref.~\cite{abdumalikov13} used pulse durations 
$\tau$ on the order of $40$ ns and transition frequencies $\nu_{ep}/(2\pi)$ on the order of 
$8$ GHz, which is well within the rotating wave approximation regime ($2\pi/(\nu_{ep} \tau)  
\approx 0.003 \ll 1$). A multiple-pulse variant of this experiment can therefore implement 
stable holonomic gates. For instance, a phase shift gate $\ket{x} \mapsto e^{ix\zeta} \ket{x}$, 
$x=0,1$, in this setup could be implemented by applying two $\pi/2$ laser pulse pairs with 
$\omega_0 = 1$, where the second pulse pair is phase shifted by $\eta = \pi - \zeta$ relative 
to the first pulse pair.

We note that the phase shift $\eta$ has only physical significance as a relative phase shift 
between the two pulse pairs. In other words, if the same phase shift had been applied in the 
original single-loop scheme of Ref.~\cite{sjoqvist12}, no physical effect would have been seen. 
In fact, the only parameters that matters for the evolution in the original scheme are the pulse 
area and the ratio $\omega_0/\omega_1$, where the latter is clearly unchanged under the 
phase shift $\omega_p \mapsto \omega_p e^{i\eta}$. 

\section{Conclusions}
Nonadiabatic holonomic quantum computation can be implemented by tailoring amplitude, 
phase, and area of laser pulses driving a $\Lambda$-systems. Here, we have proposed 
a single-loop multiple-pulse scheme that implements holonomic gates in this system. 
Specifically, we have demonstrated that the simplest nontrivial case corresponding 
to two pulse pairs ($L=2$) is sufficient to realize an arbitrary single-loop one-qubit 
gate. By combining our one-qubit gate with an entangling holonomic two-qubit 
gate, an efficient universal set of holonomic gates can be realized.  

Our scheme avoids the drawbacks of earlier versions of nonadiabatic holonomic quantum 
computation. It minimizes the exposure time to errors, but keeps the full flexibility 
concerning the choice of laser pulse shape and pulse duration. We have further outlined 
an experimental setting involving a combination of carefully chosen laser pulses. 

We note that the $L=2$ gates involve control of two new parameters: the phase shift $\eta$ 
and an additional pulse area. Thus, an optimal strategy uses the multiple-pulse one-loop 
scheme only to implement one-qubit gates with nonvanising trace (such as phase shifts); 
for gates with vanishing trace (such as bit flip and Hadamard) the original $\pi$ scheme of 
Ref.~\cite{sjoqvist12} is preferable. 

The $L=2$ case can be extended to any number of pulse pairs. The resulting paths would 
explore larger regions of the underlying Grassmann manifold $\mathcal{G} (3;2)$ and may 
therefore provide further insights into the geometrical structure of the $\mathcal{G} (3;2)$ 
holonomy. Thus, from a fundamental point of view, experimental and theoretical study of 
the $L\geq 3$ case is of interest. 
\section*{Acknowledgment} 
E. S. acknowledges support from the Swedish Research Council through Grant No. D0413201. 

\begin{eqnarray}
\mathcal{U}_1 (a_1,0) & = & \ket{d} \bra{d} + \cos a_1 \left( \hat{1} - \ket{d} \bra{d} \right) 
\nonumber \\ 
 & & -i\sin a_1 \left( \ket{b} \bra{e} + \ket{e} \bra{b} \right) , 
\nonumber \\ 
\mathcal{U}_2 (a_2,0) & = & \ket{d} \bra{d} + \cos a_2 \left( \hat{1} - \ket{d} \bra{d} \right) 
\nonumber \\ 
 & & -i\sin a_2 \left( e^{-i\eta} \ket{b} \bra{e} + e^{i\eta} \ket{e} \bra{b} \right) . 
\end{eqnarray}
The combined pulse pairs correspond to a loop provided 
\begin{eqnarray} 
 & \mathcal{U}_{2}\left( a_2,0\right)\,\mathcal{U}_{1}\left( a_1,0\right) 
\mathbb{P}(0) = \ket{d} \bra{d} 
\nonumber \\ 
 & + \left( \cos a_1 \cos a_2 - e^{-i\eta} \sin a_1 \sin a_2 \right) \ket{b} \bra{b}  
\end{eqnarray}
is unitary. This is the case if  
\begin{eqnarray}
1 & = & \left| \cos a_1 \cos a_2 - e^{-i\eta} \sin a_1 \sin a_2 \right|^2 
\nonumber \\
  & = &\sin^2 \left( \frac{\eta}{2}\right) \cos^2 (a_1 - a_2) 
\nonumber \\ 
 & & + \cos^2 \left( \frac{\eta}{2}\right) \cos^2 (a_1 + a_2) . 
\end{eqnarray}
Thus, $a_1$ and $a_2$ must be integer multiples of $\pi/2$ such that $a_1 - a_2$ and 
$a_1 + a_2$ are integer multiples of $\pi$.


\begin{thebibliography}{99}
\bibitem{remark1} $\mathcal{H}_{\Delta} (t)$ can be obtained from the original Hamiltonian 
$\mathcal{H} (t)$ in Eq.~(\ref{eq:hamiltonian1_01}), by putting $\Upsilon_p = 
\Omega (t) \omega_p$, $\Delta_0 = \Delta_1 \equiv -\Delta$, and adding a trivial 
zero-point energy term $\Delta \hat{1}$.
\bibitem{remark2} We demonstrate explicitly why $\pi/2$ pulses are essential for the two 
path segments to form a loop in the Grassmannian. To this end, we evaluate the time evolution 
operators $\mathcal{U}_1 (a_1,0)$ and $\mathcal{U}_2 (a_2,0)$ for the two sets of initial 
vectors $\ket{\psi_{1,k} (0)}$ and $\ket{\psi_{2,k} (0)}$, respectively, in the $L=2$ setting. 
We obtain
\end{thebibliography}
\end{document}